\documentclass[final,5p,times,twocolumn]{elsarticle}

\usepackage[utf8]{inputenc}
\usepackage{pdfpages}
\usepackage[hidelinks]{hyperref}
\usepackage{url}
\usepackage{amsthm,mathtools}
\usepackage{thmtools}
\usepackage{amssymb}
\usepackage{amsmath}
\usepackage{float}
\usepackage{pdflscape}
\usepackage{natbib}
\usepackage{todonotes}
\usepackage{enumitem}
\usepackage[draft]{changes}
\usepackage{dsfont}
\usepackage{graphicx}

\usepackage{tikz}
\usetikzlibrary{shapes.geometric}

\usepackage[ruled,vlined,noend]{algorithm2e}
\SetKwProg{Fn}{Function}{}{}
\DontPrintSemicolon
\SetKwFor{For}{for}{}{}
\SetKwFor{Forall}{for all}{}{}
\SetKwFor{While}{while}{}{}

\newtheorem{theorem}{Theorem}
\newtheorem{corollary}{Corollary}
\newtheorem{lemma}{Lemma}

\newtheorem{remark}{Remark}

\newcommand{\R}{\mathbb{R}}
\newcommand{\N}{\mathbb{N}}
\newcommand{\Z}{\mathbb{Z}}

\begin{document}

\begin{frontmatter}
    \title{On the Computational Complexity of Multi-Objective Ordinal \\ Unconstrained Combinatorial Optimization}

    \author[CEGIST]{Jos\'e Rui Figueira\corref{cor1}} \ead{figueira@tecnico.ulisboa.pt}
    \author[Wuppertal]{Kathrin Klamroth} \ead{klamroth@uni-wuppertal.de}
    \author[Wuppertal]{Michael Stiglmayr} \ead{stiglmayr@uni-wuppertal.de}
    \author[Wuppertal]{Julia Sudhoff Santos} \ead{sudhoff@uni-wuppertal.de}

    \address[CEGIST]{CEGIST, Instituto Superior T\'{e}cnico,  Universidade de Lisboa, Portugal}
    \address[Wuppertal]{University of Wuppertal, %
    School of Mathematics and Natural Sciences, Germany}
    \cortext[cor1]{Corresponding author.} 

    \begin{abstract}
    Multi-objective unconstrained combinatorial optimization problems (MUCO) are in general hard to solve, i.e., the corresponding decision problem is NP-hard and the outcome set is intractable. In this paper we explore special cases of MUCO problems that are actually easy, i.e., solvable in polynomial time. More precisely, we show that MUCO problems with up to two ordinal objective functions plus one real-valued objective function are tractable, and that their complete nondominated set can be computed  in polynomial time. For MUCO problems with one ordinal and a second ordinal or real-valued objective function we present an even more efficient algorithm that applies a greedy strategy multiple times. 
    \end{abstract}
    \vspace{0.25cm}
\end{frontmatter}

\section{Introduction}
\noindent Multi-objective optimization problems in general, and multi-objective combinatorial optimization (MOCO) problems in particular, have been widely studied. For a general introduction see, for example, \cite{Ehrgott2005Multicriteria} and \cite{ehrgott00annotated}. When moving from one to two or more objective functions, the computational complexity of combinatorial optimization problems usually increases dramatically. Indeed, even when a single-objective problem like, e.g., the shortest path problem, the minimum spanning tree problem, or the assignment problem is polynomially solvable, its multi-objective counterpart is usually intractable already in the bi-objective case and the associated decision problems are NP-hard in general. 
We exemplarily refer to \cite{boekler17multiobjective,ehrgott00hard, serafini87some} and the references therein. However, there are also ``easy'' classes of MOCO problems. Several such problem classes are identified in \cite{Figueira2017Easy} and the decisive properties for ``easy'' and ``hard'' instances are analyzed.  It turns out that in general neither the type of objective functions alone, nor the combinatorial structure alone, make a MOCO problem class easy in the above sense. Indeed, even unconstrained multi-objective combinatorial optimization problems (MUCO) are intractable in general and the associated  decision problems are NP-hard, see e.g.~\cite{Ehrgott2005Multicriteria,schulze19multi}. 

In this paper, we identify easy classes of MUCO problems by slightly restricting the types of considered objective functions. We show that MUCO problems with one real-valued objective function and up to two ordinal objective functions are tractable and that they can be solved in polynomial time. Our results heavily rely on the specific structure of ordinal costs, which are based on categories (like, e.g., safe, medium safe, or unsafe) rather than quantitative cost values. For an introduction to single- and multi-objective combinatorial optimization problems with ordinal costs
see, for example,  \cite{Delort2011Committee,Klamroth2023Multi, Klamroth2023Ordinal,OMahony2013Sorted,Schaefer2021binary,Schaefer2020Shortest}.

In particular, we show that the complete nondominated set of MUCO problems with one real-valued objective function and up to two ordinal objective functions can be determined by solving a polynomial number of $\varepsilon$-constraint scalarizations given by integer linear programming (ILP) problems with totally unimodular constraint matrices. Hence, such an $\varepsilon$-constraint scalarization can be solved in polynomial time by considering its linear programming relaxation. This yields an overall polynomial complexity for this relevant and non-trivial problem class. 
We present an even more efficient solution strategy for the special case of MUCO problems with one ordinal objective function and either one additional ordinal objective function or one additional real-valued objective function. 
This specific method is based on the multiple application of a greedy strategy.

The remainder of this paper is organized as follows. In Section~\ref{sec:preliminaries} we formally introduce MUCO problems and ordinal objective functions. Moreover, we recall properties of totally unimodular matrices and results from matroid theory, as we need these to prove the correctness of the algorithms presented in Section~\ref{sec:results}. We conclude the paper with an outlook in Section~\ref{sec:conclusion}.
\section{Preliminaries}\label{sec:preliminaries}
\noindent We formally introduce MUCO problems, the concept of Pareto optimality and $\varepsilon$-constraint scalarizations in  Subsection~\ref{subsec:MUCO}. Ordinal objective functions and ordinal optimality concepts are then discussed in Subsection~\ref{subsec:ordinal}. 
Subsection~\ref{subsec:OMUCO} introduces several variants of multi-objective ordinal unconstrained---and cardinality constrained---combinatorial optimization problems and their $\varepsilon$-constraint scalarizations.
We recall basic properties of totally unimodular matrices in Subsection~\ref{subsec:TU}. As transforming one ordinal objective function into constraints leads to problems with a matroid structure, we recall basic properties of matroids in Subsection~\ref{subsec:matroids}.

\subsection{Multi-objective Unconstrained Combinatorial Optimization}\label{subsec:MUCO}
\noindent We consider the \emph{multi-objective unconstrained combinatorial optimization problem~\eqref{eq:MUCO}} \begin{equation}\label{eq:MUCO}\tag{MUCO}
	\begin{array}{rl}
		\min & \bigl(z_1(x),\ldots, z_p(x) \bigr)^\top\\
		\text{s.\,t.} & x\in \{0,1\}^n
	\end{array}
\end{equation}
with $p$ linear objective functions $z_i:\{0,1\}^n\to\R$ for $i=1,\ldots,p$. If this problem has exactly two objective functions, i.e., if $p=2$, we call it \emph{bi-objective unconstrained combinatorial optimization problem} (BUCO). In the following, we consider the general case with $p\in\N$ objective functions.

We say  that an outcome vector $z(\hat{x})$ dominates an outcome vector $z(\bar{x})$, denoted by $z(\hat{x})\leq z(\bar{x})$, if it holds that $z_i(\hat{x})\leqslant z_i(\bar{x})$ for all $i=1,\ldots,p$ and $z(\hat{x})\neq z(\bar{x})$. 
If the two vectors may also be equal, we write $z(\hat{x})\leqq z(\bar{x})$.
A solution $x^*\in\{0,1\}^n$ is called \emph{efficient} if there is no other solution $x\in\{0,1\}^n$ such that $z(x)$ dominates $z(x^*)$. Then, the corresponding outcome vector $z(x^*)$ is called \emph{nondominated}. We denote by $X_E$ the set of all efficient solutions and by $Z_N$ the set of all nondominated outcome vectors. 
We aim at computing the complete nondominated set $Z_N$ and one efficient solution for each nondominated outcome vector, i.e., a minimal complete set $\bar{X}_E$ of the efficient set $X_E$. 

In general, MUCO problems are intractable (even for $p=2$) and the corresponding decision problems are NP-hard  \cite{boekler17multiobjective, serafini87some}.

One possibility to solve multi-objective optimization problems is by solving a sequence of associated single-objective optimization problems which are called scalarizations. We consider here the $\varepsilon$-constraint scalarization that transforms all but one objective functions into constraints with right-hand-side values $\varepsilon\in\R^p$:
\begin{equation}\label{eq:eMUCO}\tag{eMUCO}
	\begin{array}{rl}
		\min & z_i(x)\\
		\text{s.\,t.} & z_j(x)\leqslant \varepsilon_j \quad\text{for all} \;j\in\{1,\ldots,p\}\setminus\{i\}\\
 & x\in \{0,1\}^n.
	\end{array}
\end{equation}
It is a well-known result that every efficient solution of~\eqref{eq:MUCO} can be determined by solving~\eqref{eq:eMUCO} for an appropriate choice of $i\in\{1,\ldots,p\}$ and $\varepsilon\in\R^p$, see, e.g., \cite{Ehrgott2005Multicriteria}.

\subsection{Ordinal Costs}\label{subsec:ordinal}
\noindent Ordinal costs are used when a quantitative assessment of the quality of a certain solution element is not available. For example, when planning a bicycle path from $A$ to $B$, then, in addition to the route length, we may distinguish between safe segments (e.g., streets reserved for cyclists), medium safe segments (e.g., streets with a bicycle path), and unsafe streets (e.g., busy roads with no bicycle path). While safe is better than medium safe and medium safe is better than unsafe, no quantitative values can be assigned to these ordered categories. 
 
In general, we assume that $K$ ordered categories $\eta_1,\ldots,\eta_K$ are given, where category \(\eta_i\) is strictly preferred over category \(\eta_j\), denoted by $\eta_i\prec\eta_j$, whenever $i<j$. 
We write $\eta_i\preceq\eta_j$ if $i\leqslant j$. 
In the context of minimization problems the best category is the first one (i.e., $\eta_1$), while in the context of maximization problems the best category is the last one (i.e., $\eta_K$). 

In the context of \eqref{eq:MUCO}, we assume that every binary variable $x_i$, $i=1,\ldots,n$, represents the choice to either select an element  for a solution (i.e., $x_i=1$), or not (i.e., $x_i=0$). Let $o:\{1,\ldots,n\} \to \{\eta_1,\ldots,\eta_K\}$ denote the function that associates one of the $K$ categories with every of the $n$ elements that can be selected for a solution. 
Then we can define an associated multi-objective binary cost vector $c^i\in\{0,1\}^K$ for each element $i=1,\ldots,n$ by setting
$$c^i_j=\begin{cases}
    1,\text{ if } \eta_j \preceq o(i),\\
    0,\text{ if } \eta_j \succ o(i),\\
\end{cases}\quad j=1,\ldots,K.$$
Note that this definition implies that the binary cost vectors $c^i$ always satisfy $c^i_1=1$, and that whenever $c^i_j=0$ for some $j\in\{2,\ldots,n\}$, then also $c^i_k=0$ for all $k>j$.

Moreover, we define an incremental counting vector $c(x)\in\R^K$ as $c(x) \coloneqq c^1\,x_1+\ldots +c^n\,x_n = C\,x$ %
with $C\coloneqq(c^1,\ldots,c^n)\in\{0,1\}^{K\times n}$. 
For a solution vector $x\in\{0,1\}^n$, the outcome vector $c(x)$ hence contains in its $j$-th component the number of elements of $x$ that are in category $j$ or worse (considering minimization problems). As a consequence, $c_{j+1}(x)\leqslant c_j(x)$ for all $j=1,\ldots,K-1$. We then aim at finding solutions $x^*$ that are efficient w.r.t.\ their incremental counting vectors (ordinally efficient for short), i.e., such that there is no other solution $\hat{x}$ with $c(\hat{x})\leq c(x^*)$.
Note that a maximization problem with incremental counting vector $c(x)$ can be reformulated as a minimization problem with costs $-c(x)$. A detailed introduction to ordinal optimization problems, their specific ordering cones and the interrelation with multi-objective optimization is given, e.g., in \cite{Klamroth2023Multi}.

To illustrate these concepts,  we consider a small example instance with $n=6$ elements and $K=3$ categories.  Table~\ref{tab:ord} specifies the respective category for all six elements.
 \begin{table}[]
     \centering
     \begin{tabular}{c|cccccc}
          $i$ & $1$ & $2$ & $3$ & $4$ & $5$ & $6$  \\ \hline
          $o(i)$ & $\eta_3$ & $\eta_3$ & $\eta_1$ & $\eta_2$ & $\eta_3$ & $\eta_1$
     \end{tabular}
     \caption{Example of an instance with $n=6$ elements and $K=3$ ordinal categories.} 
     \label{tab:ord}
 \end{table}
 The corresponding cost matrix is given by
 \begin{equation}\label{eq:Cisinterval}
     C=\begin{pmatrix}
     1 & 1 & 1 & 1 & 1 & 1\\
     1 & 1 & 0 & 1 & 1 & 0\\
     1 & 1 & 0 & 0 & 1 & 0
 \end{pmatrix}.\end{equation} 
 Then, for example, the solution $\hat{x}=(1,1,0,0,1,0)^\top$ maps to the outcome vector $c(\hat{x})=(3,3,3)^\top$ (i.e., $3$ elements in $\eta_1$ or worse, $3$ elements in $\eta_2$ or worse, and $3$ elements in $\eta_3$, which means that all three elements are in category $\eta_3$), the solution $\bar{x}=(0,1,1,1,1,1)^\top$ maps to the  outcome vector $c(\bar{x})=(5,3,2)^\top$ and the solution $x'=(1,1,1,0,1,0)^\top$ maps to the  outcome vector $c(x')=(4,3,3)^\top$. If we consider a minimization problem, then $c(\hat{x})$ dominates $c(x')$. In a maximization problem it would be the other way around. In both cases, the solution $\bar{x}$ is incomparable to $\hat{x}$ and $x'$.

The set of all potentially possible outcome vectors for an ordinal objective function is denoted by 
\begin{equation}\label{eq:ordinaloutcomes}
U\coloneqq  \bigl\{u\in\Z^K_{\geq}:u_1\leqslant n\text{ and } u_{j+1}\leqslant u_j,\, j=1,\ldots,K-1 \bigr\}.
\end{equation} 

It obviously holds that $\vert U\vert=\mathcal{O}(n^K)$, see also \cite{Klamroth2023Multi}. Moreover, we denote the set of all possible outcome vectors for an ordinal objective function that can be attained when exactly $w$ elements are selected (with $w\in\{0,1,\ldots,n\}$) by 
\[
U_{w}\coloneqq\bigl\{u\in\Z^K_{\geq}:u_1=w\text{ and } u_{j+1}\leqslant u_j,\, j=1,\ldots,K-1\bigr\}.
\]
For the example from Table~\ref{tab:ord} and $w=3$ we get the following set of possible outcome vectors:
\begin{equation}\label{eq:U}
    U_3=\left\{\begin{pmatrix}
    3\\0\\0
\end{pmatrix},\begin{pmatrix}
    3\\1\\0
\end{pmatrix},\begin{pmatrix}
    3\\1\\1
\end{pmatrix},\begin{pmatrix}
    3\\2\\0
\end{pmatrix},\begin{pmatrix}
    3\\2\\1
\end{pmatrix},\begin{pmatrix}
    3\\2\\2
\end{pmatrix},\begin{pmatrix}
    3\\3\\0
\end{pmatrix},\begin{pmatrix}
    3\\3\\1
\end{pmatrix},\begin{pmatrix}
    3\\3\\2
\end{pmatrix},\begin{pmatrix}
    3\\3\\3
\end{pmatrix}\right\}.
\end{equation}

\subsection{Problems OMUCO and OMCOCC}\label{subsec:OMUCO}
\noindent In the following we investigate problems with up to two ordinal objective functions which may have different numbers of categories, in general, and that we may want to minimize or maximize, respectively. We refer to the incremental counting vector of the first ordinal objective by $\tilde{c}$ and to that of the second ordinal objective by $\hat{c}$. The corresponding numbers of categories are denoted by $\tilde{K}$ and $\hat{K}$, respectively. Analogously, we write $\tilde{C}$ and $\hat{C}$ as well as $\tilde{b}$ and $\hat{b}$, and the sets of possible outcome vectors are denoted by $\tilde{U}$ and $\hat{U}$, respectively.
In addition, we consider a real-valued sum objective function $f(x)=\sum_{i=1}^n f_i\cdot x_i\geqslant 0$ with non-negative cost coefficients $f\in\R_{\geqq}^n$.

Then, we can define the ordinal multi-objective unconstrained combinatorial optimization problem with parameters 
$\alpha,\beta,\gamma\in\{0,1,-1\}$ by
\begin{equation}\label{eq:ordMUCO}\tag{$\alpha\beta\gamma$-OMUCO}
	\begin{array}{rl}
		\min & \alpha\cdot\tilde{c}(x)\\
            \min & \beta\cdot\hat{c}(x)\\
            \min & \gamma\cdot f(x)\\
		\text{s.\,t.} & x\in \{0,1\}^n.
	\end{array}
\end{equation}
Note that a parameter value of zero (e.g., $\alpha=0$) indicates that the corresponding objective function is not considered in the optimization problem. Throughout this paper we assume that at least one of the parameters $\alpha,\beta,\gamma$ is equal to $1$ and that at least one other parameter is equal to $-1$. Otherwise, the objective functions are not conflicting and the problem is trivial. Indeed, if $\alpha=\beta=\gamma=1$, then the unique optimal solution is to set $x_i=0$ for all $i=1,\ldots,n$. Similarly, if $\alpha=\beta=\gamma=-1$ then the unique optimum is to set $x_i=1$ for $i=1,\ldots,n$. %
We denote the outcome vector of a solution $x\in\{0,1\}^n$ by
$$z(x)\coloneqq\begin{pmatrix}
    \alpha\cdot\tilde{c}(x)\\
    \beta\cdot\hat{c}(x)\\
    \gamma\cdot f(x)
\end{pmatrix}.$$

For $\gamma\neq 0$ the corresponding $\varepsilon$-constraint scalarizations are  given by
\begin{equation}\label{eq:le-ordMUCO}\tag{e-$\alpha\beta\gamma$-OMUCO}
	\begin{array}{rl}
            \min & \gamma\cdot f(x)\\
		\text{s.\,t.} & \begin{pmatrix}
		    \alpha\cdot\tilde{C}\\
                \beta\cdot \hat{C}
		\end{pmatrix}\, x\,\leqq\, \begin{pmatrix}
		    \tilde{b}\\
                \hat{b}
		\end{pmatrix}\\
            & x\in \{0,1\}^n.
	\end{array}
\end{equation}
Note that relevant right-hand-sides for these $\varepsilon$-constraint scalarizations are such that $\alpha\,\tilde{b}\in\tilde{U}$ and 
$\beta\,\hat{b}\in\hat{U}$, i.e., such that all potentially feasible outcome vectors of the ordinal objectives of problem \eqref{eq:ordMUCO} are covered.
Alternatively, $\varepsilon$-constraint scalarizations %
with equality constraints can be used, i.e., %
\begin{equation}\label{eq:e-ordMUCO}\tag{e$^{=}$-$\alpha\beta\gamma$-OMUCO}
	\begin{array}{rl}
            \min & \gamma\cdot f(x)\\
		\text{s.\,t.} & \begin{pmatrix}
		    \alpha\cdot\tilde{C}\\
                \beta\cdot\hat{C}
		\end{pmatrix}\, x= \begin{pmatrix}
		    \tilde{b}\\
                \hat{b}
		\end{pmatrix}\\
             & x\in \{0,1\}^n.
	\end{array}
\end{equation}
Note that every nondominated outcome vector $z(\bar{x})\in Z_N$ of problem~\eqref{eq:ordMUCO} can be determined by an appropriate choice of the right-hand-side vectors $(\tilde{b},\hat{b})^\top\in\R^{\tilde{K}+\hat{K}}$ with $\alpha\,\tilde{b}\in\tilde{U}$ and $\beta\,\hat{b}\in\hat{U}$
in problem~\eqref{eq:le-ordMUCO} or \eqref{eq:e-ordMUCO}, respectively, namely by setting $\tilde{b}\coloneqq \alpha\cdot\tilde{c}(\bar{x})$ and $\hat{b}\coloneqq \beta\cdot\hat{c}(\bar{x})$.

If $\gamma=0$, then the problems~\eqref{eq:le-ordMUCO} and ~\eqref{eq:e-ordMUCO} reduce to feasibility problems since in this case all feasible solutions have the same objective function value of zero. To avoid ambiguities in this case, we suggest to use one row of the matrix $\alpha\cdot\tilde{C}$ to define the objective function.

Note also that the total number of elements in any feasible solution $x$ is equal to $\tilde{c}_1(x)=\hat{c}_1(x)=\sum_{i=1}^n x_i$ since $\tilde{c}_1^i=\hat{c}_1^i=1$ for all $i=1,\ldots,n$. Hence, problem \eqref{eq:e-ordMUCO} can only be feasible if $|\tilde{b}_1|=|\hat{b}_1|=w$ for some constant $w\in\{0,1,\ldots,n\}$.

However, adding an additional cardinality constraint of the form $\sum_{i=1}^n x_i=w$ to problem \eqref{eq:ordMUCO} potentially modifies the resulting nondominated set. We hence define the ordinal multi-objective combinatorial optimization problem with cardinality constraint and with parameters $\alpha,\beta,\gamma\in\{0,1,-1\}$ by
\begin{equation}\label{eq:ordMCOCC}\tag{$\alpha\beta\gamma$-OMCOCC}
	\begin{array}{rl}
		\min & \alpha\cdot\tilde{c}(x)\\
            \min & \beta\cdot\hat{c}(x)\\
            \min & \gamma\cdot f(x)\\
		\text{s.\,t.} & \sum_{i=1}^n x_i=w,\\
                          & x\in \{0,1\}^n,
	\end{array}
\end{equation}
with $w\in\{0,1,\ldots,n\}$.  
Note that in this case the solution is not trivial even if $\alpha=\beta=\gamma=1$.

For $\gamma\neq 0$ a corresponding equality constrained $\varepsilon$-constraint scalarization is given by
\begin{equation}\label{eq:e-ordMCOCC}\tag{e$^{=}$-$\alpha\beta\gamma$-OMCOCC}
	\begin{array}{rl}
            \min & \gamma\cdot f(x)\\
		\text{s.\,t.} & \begin{pmatrix}
		    \alpha\cdot\tilde{C}\\
                \boldsymbol{1}^\top\\
                \beta\cdot \hat{C}
		\end{pmatrix}\, x= \begin{pmatrix}
		    \tilde{b}\\
                w\\
                \hat{b}
		\end{pmatrix}\\
            & x\in \{0,1\}^n,
	\end{array}
\end{equation}
with $\boldsymbol{1}=(1,\ 1,\ \ldots,\ 1)^\top\in\R^{n\times 1}$. In the following, we will also use the vector of zeros $\boldsymbol{0}=(0,\ 0,\ \ldots,\ 0)^\top\in\R^{n\times 1}$.

Throughout this paper we assume that the values $\tilde{K}$ and $\hat{K}$, i.e., the respective numbers of ordinal categories in the two ordinal objective functions, are arbitrary but fixed, and that they do not grow with the size of an instance.

\subsection{Total Unimodularity}\label{subsec:TU}
\noindent In the following we recall some important properties of totally unimodular matrices, which are also stated, for example, in \cite{Nemhauser1988Integer}.

An integral matrix $A\in\Z^{m\times n}$ is called \emph{totally unimodular (TU)} if every square submatrix $A'$ of $A$ has a determinant equal to $0$, $1$ or $-1$, i.e., $\det(A')\in\{0,1,-1\}$.

A matrix $A\in\{0,1\}^{m\times n}$ is called an \emph{interval matrix} if in each column the ones appear consecutively, i.e., if $a_{ij}=a_{kj}=1$ and $k>i+1$, then $a_{\ell j}=1$ for all $\ell\in\{i+1,\ldots,k-1\}$. The matrix $C$ from \eqref{eq:Cisinterval} is an example of an interval matrix. Indeed, it is easy to see that cost matrices of ordinal objective functions are special cases of  interval matrices, where in each column the interval of ones starts in the first row.

Interval matrices are always TU, also if some of the rows are multiplied by $-1$ (see again \cite{Nemhauser1988Integer}).

\subsection{Matroids}\label{subsec:matroids}
\noindent
A matroid $\mathcal{M}=(E,\mathcal{F})$ consists of a finite set $E$ and a set of independent sets $\mathcal{F}\subset 2^E$ such that the following three properties are satisfied:
\begin{enumerate}
    \item $\emptyset\in\mathcal{F}$
    \item $F\in\mathcal{F}$ and $H\subseteq F\ \implies\ H\in\mathcal{F}$
    \item $F,H\in\mathcal{F}:\ \vert H\vert <\vert F\vert\ \implies\  \exists f\in F\setminus H:\ H\cup \{f\}\in\mathcal{F}$.
\end{enumerate}
All inclusion wise maximal independent sets are called bases of the matroid. It is well-known that all bases have the same cardinality, see, e.g., \cite{Oxley1992Matroid}. A specific matroid that we use in the following is the so-called partition matroid. In this case, the set $E$ is the disjoint union of $k$ finite sets, i.e., $E=E_1\cup\ldots\cup E_k$. The set of independent sets is then defined as $\mathcal{F}\coloneqq\{F\subseteq E:\ \vert F\cap E_i\vert\leqslant u_i\ \forall i=1,\ldots,k\}$ for given upper bounds $u_1,\ldots,u_k\geqslant 0$.

Matroid Optimization Problems~\eqref{eq:MOP} are defined as
\begin{equation}\label{eq:MOP}\tag{MOP}
	\begin{array}{rl}
            \min & z(x)\\
		\text{s.\,t.} & x\in \mathcal{B},
	\end{array}
\end{equation}
with $\mathcal{B}$ the set of all bases of a matroid. Such problems can be solved with a greedy strategy, no matter whether the objective function is a real-valued sum objective function or an ordinal objective function, see \cite{Klamroth2023Multi}.

\section{Theoretical Results}\label{sec:results}

\noindent In the following we present some theoretical results regarding the nondominated and efficient set of the problems~\eqref{eq:ordMUCO} and \eqref{eq:ordMCOCC}. We first present some general results and a general solution strategy. Then we consider the specific cases of both problem types with only two objective functions.

Before we consider the general case, we note that there are some obvious efficient solutions for problem~\eqref{eq:ordMUCO}:
\begin{lemma}\label{rem:solution}
    We consider the problem~\eqref{eq:ordMUCO} and assume that not all parameters $\alpha,\ \beta,\ \gamma$ are equal to zero. Then we note the following:
    \begin{enumerate}
        \item[(a)] If $\alpha,\beta,\gamma\in\{0,1\}$ and if at least one of the parameters $\alpha,\beta$ is equal to one, then the efficient set contains only the zero vector, i.e., $X_E=\{\boldsymbol{0}\}$. 
        \item[(b)] %
        If $\alpha,\beta,\gamma\in\{0,-1\}$ and if at least one of the parameters $\alpha,\beta$ is equal to minus one, then the efficient set contains only the all ones vector, i.e., $X_E=\{\boldsymbol{1}\}$.
        \item[(c)] If there exists one of the parameters $\alpha,\beta,\gamma$ that is equal to $1$ and another one that is equal to $-1$, then it holds that $\{\boldsymbol{0},\boldsymbol{1}\}\subseteq X_E$. 
    \end{enumerate}
\end{lemma}
\begin{proof}
    This result follows immediately since we consider solely unconstrained optimization problems: If positive costs are to be minimized, then choosing no element (or item) is always the unique optimal solution. Similarly, if positive costs are to be maximized, then choosing all elements is always the unique optimal solution. If there is a trade-off between maximization and minimization, then these two extreme solutions remain efficient.
\end{proof}
Note that when $\alpha=\beta=0$ and $\gamma=1$ for an instance of problem~\eqref{eq:ordMUCO} then we always have $\{0\}\subseteq X_E$. Analogously, when when $\alpha=\beta=0$ and $\gamma=-1$ then $\{\boldsymbol{1}\}\subseteq X_E$. However, further solutions may be efficient in these cases since some of the coefficients $f_i$, $i\in\{1,\ldots,n\}$, may be equal to zero. 
Note also that the corresponding constrained variant, problem~\eqref{eq:ordMCOCC}, is non-trivial even if $\alpha,\beta,\gamma\in\{0,1\}$ or $\alpha,\beta,\gamma\in\{0,-1\}$.

To illustrate possible dominance situations, we consider two instances of problem~\eqref{eq:ordMUCO} with $\alpha=\beta=1$, $\gamma=-1$, $n=4$, and $\tilde{K}=\hat{K}=3$. The categories and the objective coefficients of the four elements in the respective instances are specified in the Tables~\ref{tab:ex1} and \ref{tab:ex2}.
 \begin{table}[]
     \centering
     \begin{tabular}{c|cccc}
          $i$ & $1$ & $2$ & $3$ & $4$  \\ \hline
          $\tilde{o}(i)$ & $\eta_2$ & $\eta_1$ & $\eta_2$ & $\eta_1$\\
          $\hat{o}(i)$ & $\eta_3$ & $\eta_2$ & $\eta_1$ & $\eta_3$\\
          $f_i$ & $10$ & $1$ & $3$ & $2$ 
     \end{tabular}
     \caption{Instance A of problem ($1,1,-1$-OMUCO).\label{tab:ex1}}
 \end{table}
  \begin{table}[]
     \centering
     \begin{tabular}{c|cccc}
          $i$ & $1$ & $2$ & $3$ & $4$  \\ \hline
          $\tilde{o}(i)$ & $\eta_1$ & $\eta_2$ & $\eta_1$ & $\eta_2$\\
          $\hat{o}(i)$ & $\eta_2$ & $\eta_1$ & $\eta_1$ & $\eta_2$\\
          $f_i$ & $10$ & $5$ & $1$ & $11$ 
     \end{tabular}
     \caption{Instance B of problem ($1,1,-1$-OMUCO).\label{tab:ex2}}
 \end{table}
The first example, instance $A$ in Table~\ref{tab:ex1}, shows that it may happen that a solution $x^1$ that selects only one element is preferred over a solution $x^2$ that selects several elements (i.e., $z(x^1)$ dominates $z(x^2)$), while $x^1$ is not preferred over any solution containing only a subset of the elements in $x^2$ (i.e., $z(x^1)$ does not dominate the outcome vector of any such solution). Indeed, if $x^1=(1,0,0,0)^\top$ and $x^2=(0,0,1,1)^\top$, then $z(x^1)=(1,1,0,1,1,1,-10)^\top\leq (2,1,0,2,1,1,-5)^\top=z(x^2)$. 
However, $x^1$ is not preferred over any solution selecting only a subset of the last two elements.

The second example, instance $B$ in Table~\ref{tab:ex2}, shows that it may happen that the individual elements are pairwise incomparable, and the outcome vector of no individual element dominates the outcome vector of any larger set of elements. However, there is a set of two elements that is preferred over another set of two elements. Indeed, for $x^3=(1,1,0,0)^\top$ and $x^4=(0,0,1,1)^\top$ we have $z(x^3)=(2,1,0,2,1,0,-15)^\top\leq (2,1,0,2,1,0,-12)^\top=z(x^4)$. All other solutions are efficient, i.e., $X_E=\{0,1\}^4\setminus \{(0,0,1,1)^\top\}$.

\subsection{Polynomial Time Algorithm for OMUCO and OMCOCC}

\noindent The following results hold for general choices of the parameters $\alpha,\beta,\gamma\in\{-1,0,1\}$.
\begin{lemma}\label{lemma:tractable}
    Problem~\eqref{eq:ordMUCO} is tractable.
\end{lemma}
\begin{proof}
    Since every nondominated outcome vector of problem \eqref{eq:ordMUCO} can be obtained by solving an $\varepsilon$-constraint scalarization \eqref{eq:e-ordMUCO} with an appropriate right-hand-side vector $(\tilde{b},\hat{b})^\top\in\R^{\tilde{K}+\hat{K}}$ with $\alpha\,\tilde{b}\in\tilde{U}$ and $\beta\,\hat{b}\in\hat{U}$, 
    the cardinality of $Z_N$ is bounded by the number of such relevant right-hand-side vectors. Towards this end, we recall the definition of the set of possible outcome vectors for an ordinal objective function given in \eqref{eq:ordinaloutcomes}. Hence, a superset of all potentially relevant right-hand-side vectors is given by 
    \begin{equation}\label{eq:Ue}
    U^e \coloneqq \left\{(\tilde{b},\hat{b})^\top\!\in\!\R^{\tilde{K}+\hat{K}}:\, \alpha\,\tilde{b}\!\in\!\tilde{U},\, \beta\,\hat{b}\!\in\!\hat{U},  \text{~and~} \bigl|\tilde{b}_1\bigr|\!=\!\bigl|\hat{b}_1\bigr|\,\right\}.
    \end{equation}
    Since $|\tilde{U}|=\mathcal{O}(n^{\tilde{K}})$ and $|\hat{U}|=\mathcal{O}(n^{\hat{K}})$ we obtain $|U^e|=\mathcal{O}(n^{\tilde{K}+\hat{K}-1})$, and the result follows.
\end{proof}

\begin{theorem}\label{thm:constraintTU}
    The constraint matrix $$A=\begin{pmatrix}
		    \alpha\cdot\tilde{C}\\
                \beta\cdot \hat{C}
		\end{pmatrix}$$ of problem~\eqref{eq:e-ordMUCO} is TU.
\end{theorem}
\begin{proof}
    First note that the rows in $A$ 
    can be rearranged without changing the feasible set of the optimization problem. We may hence rearrange the rows of the submatrix $\alpha\cdot\tilde{C}$ in reverse order,
    i.e., the last row of $\alpha\cdot\tilde{C}$ becomes its first row, the second but last row becomes its second row, and so on. The resulting matrix is denoted by $\alpha\cdot\tilde{C}'$. It is easy to see that the matrix $$A'=\begin{pmatrix}
		    \alpha\cdot\tilde{C}'\\
                \beta\cdot \hat{C}
		\end{pmatrix}$$
    is an interval matrix (with some of its rows possibly multiplied by $-1$). Then the result follows immediately as interval matrices are TU, see \cite{Nemhauser1988Integer}.
\end{proof}

\begin{corollary}\label{cor:constraintTU}
    The constraint matrix $$A=\begin{pmatrix}
		    \alpha\cdot\tilde{C}& I_{\tilde{K}} & 0\\
                \beta\cdot\hat{C}& 0 & I_{\hat{K}}
		\end{pmatrix}$$ of problem~\eqref{eq:le-ordMUCO} in standard form is TU.
\end{corollary}
\begin{proof}
    Since a matrix that is obtained from a TU matrix by appending an identity matrix is also TU (see, e.g., \cite{Nemhauser1988Integer}), this is an immediate consequence of Theorem~\ref{thm:constraintTU}.
\end{proof}

\begin{corollary}\label{cor:OMCOCC}
    The constraint matrix $$A=\begin{pmatrix}
		    \alpha\cdot\tilde{C}\\
                \boldsymbol{1}^\top\\
                \beta\cdot \hat{C}
		\end{pmatrix}$$ of problem~\eqref{eq:e-ordMCOCC} is TU.
\end{corollary}
\begin{proof}
    After rearranging the rows of the submatrix $\alpha\cdot\tilde{C}$ as in the proof of  Theorem~\ref{thm:constraintTU}, we obtain an equivalent constraint systems with constraint matrix
    $$A'=\begin{pmatrix}
		    \alpha\cdot\tilde{C}'\\
                \boldsymbol{1}^\top\\
                \beta\cdot \hat{C}.
		\end{pmatrix}$$
    Clearly, this matrix is again an interval matrix (with some rows possibly multiplied by $-1$), and hence $A$ is TU.
\end{proof}

\begin{remark}
The above results do not generalize to multi-objective unconstrained combinatorial optimization problems with three or more ordinal objective functions. Indeed, for higher dimensional problems the constraint matrix of the associated $\varepsilon$-constraint scalarizations is in general not TU.
\end{remark}

Lemma~\ref{lemma:tractable} and Theorem~\ref{thm:constraintTU} imply a simple, yet polynomial, algorithm for problem~\eqref{eq:ordMUCO}: Enumerate all of the $\mathcal{O}(n^{\tilde{K}+\hat{K}-1})$ potentially relevant right-hand-side vectors from the set $U^e$ (cf.\ \eqref{eq:Ue}), solve an associated $\varepsilon$-constraint scalarization \eqref{eq:e-ordMUCO} for each of them (or its inequality-constrained variant \eqref{eq:le-ordMUCO}), and filter out all dominated outcome vectors at the end. 
Since all vectors in $U^e$ are integral, and since the constraint matrix of problems \eqref{eq:le-ordMUCO} (in standard form) and \eqref{eq:e-ordMUCO} are TU, each $\varepsilon$-constraint scalarization can be solved in polynomial time using an appropriate LP solver for its linear programming relaxation. We refer to \cite{karmarkar84} and \cite{vaidya89} for applicable polynomial time linear programming algorithms, where the latter achieves a worst case complexity of $\mathcal{O}(n^{2.5}\,\log_2 n)$ in the special case considered here: A TU constraint matrix $A$ with coefficients from $\{-1,0,1\}$, and with a comparably small number of rows that does not grow with the size of the instance. The complete method is summarized in Algorithm~\ref{alg:generalMUCO}.

\begin{algorithm}
	\SetAlgoLined 
	\LinesNumbered
	\KwIn{Instance of problem \eqref{eq:ordMUCO}}
	\KwOut{Minimal complete set $\bar{X}_E$ of the efficient set of problem \eqref{eq:ordMUCO}}
	\caption{Exact polynomial time solution algorithm for \eqref{eq:ordMUCO}}\label{alg:generalMUCO}
        $\bar{X}\coloneqq \emptyset$\;
        \Forall{$(\tilde{b},\hat{b})^\top\in U^e$}{ 
            Solve the LP relaxation of \eqref{eq:e-ordMUCO} with right-hand-sides $(\tilde{b},\hat{b})^\top$ and save the identified optimal solution $x$ (if the problem is feasible)\;
            $\bar{X}=\bar{X}\cup \{x\}$\;
        }
        Filter $z(\bar{X})$ for dominated outcome vectors and remove the non-efficient pre-images from $\bar{X}$\;
	\Return $\bar{X}_E\coloneqq\bar{X}$
\end{algorithm}

Since the final filter operation in Step~5 of Algorithm~\ref{alg:generalMUCO} requires, in the worst case, a pairwise comparison of the set of generated outcome vectors, 
and since this set has a polynomial cardinality (as discussed above), Algorithm~\ref{alg:generalMUCO} is indeed a polynomial time algorithm.

\begin{corollary}\label{cor:polyalg}
For fixed values of $\tilde{K}$ and $\hat{K}$, the nondominated set
    $Z_N$ and a minimal complete set
    $\bar{X}_E$ of problem~\eqref{eq:ordMUCO} can be computed in polynomial time. 
\end{corollary}

\begin{remark}
The result of Corollary~\ref{cor:polyalg} immediately transfers to  problem~\eqref{eq:ordMCOCC}. In this case, the set of relevant outcome vectors $U^e$ as defined in \eqref{eq:Ue} needs to be replaced by its subsets representing solutions with exactly $w$ elements, i.e., 
\begin{equation*}%
    U^e_w \coloneqq \left\{(\tilde{b},\hat{b})^\top\!\in\!\R^{\tilde{K}+\hat{K}}:\, \alpha\,\tilde{b}\!\in\!\tilde{U},\, \beta\,\hat{b}\!\in\!\hat{U},  \text{~and~} \bigl|\tilde{b}_1\bigr|\!=\!\bigl|\hat{b}_1\bigr|\!=\!w\,\right\}
    \end{equation*}
with $w\in\{0,1,\ldots,n\}$. Note that $U^e=\bigcup_{w=0,1,\ldots,n}U^e_w$. For a fixed value of $w$, the set $U^e$ in Line~2 of Algorithm~\ref{alg:generalMUCO} needs to be replaced by the correct subset $U^e_w$. Moreover, for each $(\tilde{b},\hat{b})^\top\in U^e_w$, the $\varepsilon$-constraint scalarization \eqref{eq:e-ordMCOCC} has to be solved in Line~3 of the algorithm, with right-hand-side vector $(\tilde{b},w,\hat{b})^\top$. That these subproblems are also TU follows from Corollary~\ref{cor:OMCOCC}.
\end{remark}

\begin{remark}
Algorithm~\ref{alg:generalMUCO} can be modified so that the filter operation in Line~5 can be avoided. Indeed, if the $\varepsilon$-constraint scalarizations with equality constraints \eqref{eq:e-ordMUCO} are replaced by augmented $\varepsilon$-constraint scalarizations of the form
\begin{equation*}%
	\begin{array}{rl}
            \min & \displaystyle\gamma \,f(x) + \delta \left(\alpha\sum\limits_{j=1}^{\tilde{K}}\tilde{c}_j(x) + \beta\sum\limits_{j=1}^{\hat{K}}\hat{c}_j(x) \right)\\[2.75ex]
		\text{s.\,t.} & \begin{pmatrix}
		    \alpha\cdot\tilde{C}\\
                \beta\cdot\hat{C}
		\end{pmatrix}\, x \,\leqq\, \begin{pmatrix}
		    \tilde{b}\\
                \hat{b}
		\end{pmatrix}\\
             & x\in \{0,1\}^n
	\end{array}
\end{equation*}
with an augmentation parameter $\delta>0$, then all generated solutions are guaranteed to be efficient (and not just weakly efficient as in the case of problem \eqref{eq:e-ordMUCO}) \cite{mavrotas2009}. To ensure that no non-dominated point is missed, the value of $\delta$ should be chosen such that
\[
\delta< \frac{1}{(\tilde{K}\!+\!\hat{K})\cdot n}.
\]
Moreover, in this case irrelevant right-hand-side vectors $(\tilde{b},\hat{b})^\top\in U^e$ can be avoided by selecting the subproblems (i.e., relevant right-hand-sides $(\tilde{b},\hat{b})^\top$) based on an objective space method, see, e.g., \cite{daechert2024}.
\end{remark}

\subsection{Greedy-based Algorithm for the Bi-objective Case}

\noindent In this section we focus on bi-objective variants of the problems~\eqref{eq:ordMUCO} and \eqref{eq:ordMCOCC}, i.e., we assume that $\alpha=0$ and $\beta\cdot \gamma=-1$, $\beta=0$ and $\alpha\cdot\gamma=-1$ or $\gamma=0$ and $\alpha\cdot\beta=-1$. Note that in all these cases, at least one of the objective functions is ordinal.

In the following we assume w.l.o.g.\ that either $\beta=0$ or $\gamma=0$. For problem~\eqref{eq:ordMUCO}, we consider $\varepsilon$-constraint scalarizations with equality constraints for the ordinal objective function $\alpha\,\tilde{c}(x)$, i.e.,
\begin{equation}\label{eq:e-ordMUCO2}\tag{e$^=$-$\alpha\beta\gamma$-OMUCO-2}
	\begin{array}{rl}
            \min & \begin{cases}
                \gamma\cdot f(x) & \text{if}\; \gamma\neq 0\\
                \beta\cdot \hat{C}\, x & \text{if}\; \gamma =0 
            \end{cases}\\
		\text{s.\,t.} & 
		    \alpha\cdot\tilde{C}\, x = 
		    \tilde{b}\\
             & x\in \{0,1\}^n,
	\end{array}
\end{equation}
with right-hand-sides $\tilde{b}$ such that $\alpha\cdot\tilde{b}\in\tilde{U}$.

When considering problem~\eqref{eq:ordMCOCC}, then the $\varepsilon$-constraint scalarizations can be formulated analogously, however, with the additional constraint $\boldsymbol{1}^\top\,x=w$ (or, equivalently, by setting $|\tilde{b}_1|=w$) for some constant $w\in\{0,1,\ldots,n\}$. 

In the following we will interrelate the feasible set of the $\varepsilon$-constraint scalarization \eqref{eq:e-ordMUCO2} with an associated partition matroid. This is the basis for solving problem \eqref{eq:e-ordMUCO2} with a greedy algorithm.

\begin{theorem}\label{th:greedy}
    If the $\varepsilon$-constraint problem \eqref{eq:e-ordMUCO2} of the biobjective version of problem \eqref{eq:ordMUCO} is feasible, then an optimal solution can be computed using a greedy algorithm.
\end{theorem}
\begin{proof}
  Suppose that a given instance of problem~\eqref{eq:e-ordMUCO2} with right-hand-side vector $\tilde{b}\in\R^{\tilde{K}}$ is feasible. We now consider an associated partition matroid $\mathcal{M}=(E,\mathcal{F})$ with ground set $E=\{1,\ldots,n\}$.
  The ground set $E$ is partitioned into $\tilde{K}$ pairwise disjoint sets $E_i$, $i=1,\ldots,\tilde{K}$ such that $E_i$ contains exactly those elements that are in category $\eta_i$,  i.e., $E_i\coloneqq \{j\in E\,:\, \tilde{o}(j)=\eta_i\}$ %
  and $E=E_1\cup\cdots\cup E_{\tilde{K}}$. The independent sets of $\mathcal{M}$ are defined as $\mathcal{F}\coloneqq\{F\subseteq E:\ |F\cap E_i|\leqslant |\tilde{b}_i|-|\tilde{b}_{i+1}|,\, i=1,\ldots,\tilde{K}\}$. For simplicity of notation we define $\tilde{b}_{\tilde{K}+1}\coloneqq 0$. Note that this implies that $|F|\leqslant |\tilde{b}_{1}|-|\tilde{b}_{2}|+\cdots+ |\tilde{b}_{\tilde{K}-1}|-|\tilde{b}_{\tilde{K}}|+|\tilde{b}_{\tilde{K}}|-|\tilde{b}_{\tilde{K}+1}|=|\tilde{b}_1|$ for all $F\in\mathcal{F}$.

  Since we have assumed that problem~\eqref{eq:e-ordMUCO2} is feasible, it holds that $|E_i|\geqslant |\tilde{b}_i|-|\tilde{b}_{i+1}|$ for all $i=1,\ldots,\tilde{K}$. This implies that all bases $B\in\mathcal{B}$ of $\mathcal{M}$ satisfy $|B\cap E_i|=|\tilde{b}_i|-|\tilde{b}_{i+1}|$, $i=1,\ldots,\tilde{K}$, and hence have cardinality $|B|=|\tilde{b}_1|$. Finally, we can observe that there is a one-to-one correspondence between feasible solution vectors $x\in\{0,1\}^n$ of problem~\eqref{eq:e-ordMUCO2} and bases $B\in\mathcal{B}$ of this partition matroid by setting $x_i=1$ if and only if $i\in B$. 
  Since matroid optimization problems with a real-valued or with an ordinal objective function can be solved by a greedy algorithm, see \cite{Klamroth2023Multi}, the result follows.
\end{proof}

Note that the feasibility of the $\varepsilon$-constraint scalarization \eqref{eq:e-ordMUCO2} for a particular right-hand-side vector $\tilde{b}\in\R^{\tilde{K}}$ with $\alpha\,\tilde{b}\in\tilde{U}$
can be easily checked using the condition from the proof of Theorem~\ref{th:greedy}: Problem~\eqref{eq:e-ordMUCO2} is feasible if and only if $|E_i|\geqslant |\tilde{b}_i|-|\tilde{b}_{i+1}|$ for all $i=1,\ldots,\tilde{K}$ (where we set again $\tilde{b}_{\tilde{K}+1}\coloneqq 0$). Note also that the $\varepsilon$-constraint scalarization of the biobjective version of problem~\eqref{eq:ordMCOCC}, that has the additional constraint $\boldsymbol{1}^\top\,x=w$, is feasible for a particular right-hand-side vector $(\tilde{b},w)^\top\in\R^{\tilde{K}+1}$ if and only if problem~\eqref{eq:e-ordMUCO2} is feasible for $\tilde{b}$ and  $w=|\tilde{b}_1|$.

\begin{corollary}
    If the $\varepsilon$-constraint problem \eqref{eq:e-ordMUCO2} with the additional constraint $\boldsymbol{1}^\top\,x=w$ of the biobjective version of problem \eqref{eq:ordMCOCC} is feasible, then an optimal solution can be computed using a greedy algorithm.
\end{corollary}

Overall, these results lead to Algorithm~\ref{alg:MUCO2} that solves problem~\eqref{eq:ordMUCO} %
with two conflicting objective functions in time $\mathcal{O}(n\, \log(n)+n^{\tilde{K}+1}+n^{2\,\tilde{K}})= \mathcal{O}(n^{2\,\tilde{K}})$.%

\begin{algorithm}
	\SetAlgoLined 
	\LinesNumbered
	\KwIn{Instance of problem~\eqref{eq:ordMUCO} with two conflicting objective functions}
	\KwOut{Minimal complete set $\bar{X}_E$ of the efficient set of problem \eqref{eq:ordMUCO}} 
	\caption{Greedy-based algorithm for \eqref{eq:ordMUCO} with two objective functions}\label{alg:MUCO2}
        $\bar{X}\coloneqq \emptyset$, $\tilde{b}_{\tilde{K}+1}\coloneqq 0$\;
        Partition the elements/their indices into $\tilde{K}$ sets $E_i$, $i=1,\ldots,\tilde{K}$, such that $E_i$ contains all elements from category $\eta_i$, i.e., $E_i\coloneqq \{j\in \{1,\ldots,n\}\,:\, \tilde{o}(j) = \eta_i\}$ %
        \;
        Sort the elements in all sets $E_i$ such that they are non-improving with respect to the second objective\;
        \Forall{$\alpha\cdot \tilde{b}\in \tilde{U}$}{
            $x\coloneqq \boldsymbol{0}$\;
            \For{$i=1,\ldots,\tilde{K}$}{
                \If{$\vert E_i\vert \geqslant |\tilde{b}_i|-|\tilde{b}_{i+1}|$}{
                    Choose the first $(|\tilde{b}_i|-|\tilde{b}_{i+1}|)$ elements in $E_i$ and set the corresponding variables in $x$ to $1$\;
                }\Else{
                    Break\;
                }
            }
            \If{$\sum_{i=1}^n x_i=\tilde{b}_1$}{ 
            $\bar{X}=\bar{X}\cup \{x\}$\;
            }
        }
        Filter $z(\bar{X})$ for dominated outcome vectors and remove the non-efficient pre-images from $\bar{X}$\;
	\Return $\bar{X}_E\coloneqq \bar{X}$
\end{algorithm}

Algorithm~\ref{alg:MUCO2} can be easily adapted to solve a biobjective instance of problem \eqref{eq:ordMCOCC}, i.e., problem \eqref{eq:ordMUCO} with the additional constraint $\boldsymbol{1}^\top\,x=w$ with $w\in\{0,1,\ldots,n\}$. The only necessary change occurs in line~4, where the set $\tilde{U}$ has to be replaced by the set $\tilde{U}_w$.

To illustrate Algorithm~\ref{alg:MUCO2}, we consider again the example instance introduced in Table~\ref{tab:ord} and assume that the ordinal objective function with cost matrix $\tilde{C}\coloneqq C$ as specified in \eqref{eq:Cisinterval} is to be maximized, i.e., $\alpha=-1$. Suppose that an additional real-valued sum objective function is given by $f(x)=\sum_{i=1}^6 f_ix_i$ with coefficients $f_i=i$, $i=1\ldots,6$. Let $\gamma=1$, i.e., $f$ is to be minimized.  We want to solve the corresponding problem~\eqref{eq:ordMCOCC} with $\beta=0$ and $w=3$.  Algorithm~\ref{alg:MUCO2} first identifies and sorts the sets $E_1=\{3,6\}$, $E_2=\{4\}$ and $E_3=\{1,2,5\}$. Then, all elements of the set $\tilde{U}_3\coloneqq U_3$ (cf.\ equation~\eqref{eq:U}) are considered. The vector $\tilde{b}=(-3,0,0)^\top$ is discarded since  $|E_1|=2<3=|\tilde{b}_1|-|\tilde{b}_2|$. For $\tilde{b}=(-3,-1,0)^\top$ the greedy algorithm returns the solution $x^1=(0,0,1,1,0,1)^\top$ with $\gamma f(x^1)=13$ and $\alpha \tilde{C} x^1=(-3,-1,0)^\top$. Then, for all vectors 
$$\tilde{b}\in\left\{\begin{pmatrix}
    -3\\-2\\0
\end{pmatrix},\begin{pmatrix}
    -3\\-3\\0
\end{pmatrix},\begin{pmatrix}
    -3\\-3\\-1
\end{pmatrix}\right\}$$ 
there is again no solution since $|E_2|=1$. Table~\ref{tab:sol} 
provides a complete list of all solutions that are generated during the course of  Algorithm~\ref{alg:MUCO2}.  Finally, after the filter operation in line~13 of Algorithm~\ref{alg:MUCO2} the solutions $x^1$, $x^2$ and $x^3$ are discarded as their outcome vectors $z(x^1)=-\tilde{C} x^1$, $z(x^2)=-\tilde{C} x^2$ and $z(x^3)=-\tilde{C} x^3$ are dominated by $z(x^4)=-\tilde{C} x^4$. Thus, Algorithm~\ref{alg:MUCO2} returns $\bar{X}_E=\{x^4,x^5,x^6\}$.
\begin{table}
     \centering
     \begin{tabular}{c|c|c}
          $x^i$ & $f(x^i)$ & $-\tilde{C} x^i$ \\ \hline
          $x^1=(0,0,1,1,0,1)^\top$ & $13$ & $(-3,-1,0)^\top$\\
          $x^2=(1,0,1,0,0,1)^\top$ & $10$ & $(-3,-1,-1)^\top$\\
          $x^3=(1,0,1,1,0,0)^\top$ & $8$ & $(-3,-2,-1)^\top$\\
          $x^4=(1,1,1,0,0,0)^\top$ & $6$ & $(-3,-2,-2)^\top$\\
          $x^5=(1,1,0,1,0,0)^\top$ & $7$ & $(-3,-3,-2)^\top$\\
          $x^6=(1,1,0,0,1,0)^\top$ & $8$ & $(-3,-3,-3)^\top$\\
     \end{tabular}
     \caption{Greedy solutions returned by Algorithm~\ref{alg:MUCO2} for the problem~\eqref{eq:ordMCOCC} with $\alpha=-1$, $\beta=0$, $\gamma=1$, and $w=3$ for the example instance specified in Table~\ref{tab:ord}. The real-valued objective function has coefficients $f_i=i$, $i=1,\ldots,6$.\label{tab:sol}}
     \label{tab:ex4}
 \end{table}

\section{Conclusion}\label{sec:conclusion}
\noindent MUCO problems are in general intractable and the corresponding decision problems are usually NP-hard. Despite this general difficulty, we show in this paper that the specific case of MUCO problems with one real-valued sum objective  and two ordinal objectives are tractable and solvable in polynomial time.
Our results are based on the formulation of a series of associated $\varepsilon$-constraint scalarizations that convert the two ordinal objective functions into constraints. Since these $\varepsilon$-constraint scalarizations have totally unimodular constraint matrices, they can be solved by linear programming. Moreover, the outcome values of the ordinal objective functions are polynomially bounded, and hence the number of potentially nondominated outcome vectors of ordinal MUCO problems is also polynomially bounded. Overall, we obtain a polynomial time solution algorithm for this special class of MUCO problems.

For the biobjective case, i.e., for MUCO problems with two ordinal objectives, or with one ordinal and one real-valued sum objective, we derive an even more efficient solution method that is based on the repeated application of a simple greedy algorithm. The correctness of this approach is based on an interrelation between the associated $\varepsilon$-constraint scalarizations and partition matroids, for which optimal solutions can be obtained very efficiently.

Future research should focus on further properties of MUCO problems with ordinal objective functions. An important open question is whether -- or under what conditions -- the nondominated set of ordinal MUCO problems is connected. 
Moreover, problems with many (ordinal) objective functions and problems with a more specific combinatorial structure could be analyzed.

\section*{Acknowledgements}
\addcontentsline{toc}{section}{\numberline{}Acknowledgements}
\noindent José Rui Figueira acknowledges the support by national funds through FCT (Fundação para a Ciência e a Tecnologia), Portugal under the project UIDB/00097/2020. The authors acknowledge the financial support of the DAAD/FCT project OCO -- Ordinal Combinatorial Optimization (DAAD-ID: 57711909).

\end{document}